\documentclass[runningheads]{llncs}
\usepackage[T1]{fontenc}
\usepackage{graphicx}
\usepackage{url}
\usepackage{color}
\begin{document}
\title{The International Lunar Reference System}
%
%
\author{Agnès  Fienga\inst{1}\orcidID{0000-0002-4755-7637} on behalf of the IAG JWG 1.1.3 }


%
\authorrunning{A. Fienga et al.}
%
\institute{(1) Observatoire de la Côte d'azur, University Côte d'Azur, CNRS, 250 av. A. Einstein , Valbonne, France \email{agnes.fienga@oca.eu}\\
}
\maketitle              
\begin{abstract}
As the exploration of the Moon accelerates in the coming years, there is already an urgent need to standardise the Position, Navigation and Timing of spacecraft, and consequently for the definition of the lunar reference system and frame. This document gathers the most recent recommendations of the the International Association of Geodesy (IAG) / International Astronomical Union (IAU)  Joint Working Group (WG) 1.1.3\textcolor{black}{ "Lunar Reference frames"} on the topic. Despite the challenges for producing a Lunar Reference System and Frame due to the limited number of control points and a single data type, the WG proposes applying an approach developed for Global Navigation Satellite Systems (GNSS). This supports an accurate and reliable reference system, consistent with International Earth Rotation and Reference Systems Service (IERS) standards and, based on redundant sources, ensures resilience for the lunar Position, Navigation and Timing  (PNT) framework. The first realization of the International Lunar Reference Frame 2026 (ILuRF2026), ILuRF2026, is delivered on the temporay website \url{https://ilurs-6e772d.gitlab.io/} and on the future \url{http://ilurf.gssc.esa.int} together with associated software.

\end{abstract}
%
%
%
\section{Introduction}
Since 2023, \textcolor{black}{a significant number of} missions to the Moon are planned, including 6 to 10 crewed ones. Because of potential hazard to human life, the need for accurate localization on the lunar surface is increasing. Various actors will also be involved during this intense phase of exploration, highlighting the need for standardisation of procedures among space agencies and private companies. Since 2025, the United Nations Office for Outer Space Affairs organises the annual CisLunar PNT workshop, bringing together scientists and engineers to address these issues and since 2022, the IAG and IAU established the IAG/ IAU Joint WG 1.1.3 \textcolor{black}{"Lunar Reference frames"} for discussing the questions related to the definition of the lunar reference system, keystone for accurate and standardized PNT procedures. 
This document gathers the most recent recommendations of the WG 1.1.3 concerning the definition of a standardized International Lunar Reference System.

\begin{figure}
\centering
\includegraphics[width=0.95\textwidth]{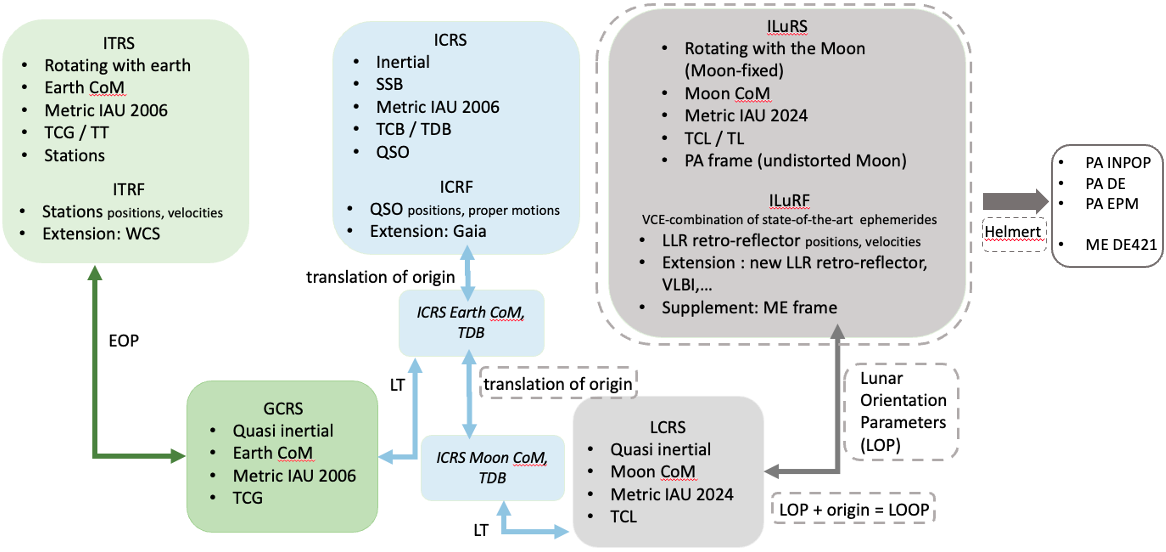}
\caption{Transformations from the \textcolor{black}{International Celestial Reference System (ICRS), International Terrestrial Reference System (ITRS), Lunar Celestial Reference System (LCRS) and the International Lunar Reference System (ILuRS)}. The dashed zones correspond to new definition and transformation proposed by the WG 1.1.3.  \textcolor{black}{'LT' flags a required Lorentz transformation due to differing metrics between reference systems. The ICRF-to-LCRS case, for example, involves two steps: i) origin translation from SSB to the Moon's center of mass, and ii) metric change (LT) from SSB to Moon, affecting space and time.} } 
\label{fig:ILuRS_workflow}
\end{figure}

\section{International Lunar Reference System}
\label{sec:ILuRS}
The International Lunar Reference System (ILuRS) follows the criteria of \cite{Kovalevsky81} (see Fig.~\ref{fig:ILuRS_workflow}).  
\textcolor{black}{Two possible structures can be chosen:  the Mean-Earth frame (ME) based on the averaged Earth-Moon orientation (with the x-axis pointing toward the Earth-mean direction seen from the Moon and the z-axis is along the mean rotational pole) and the Principal Axis of the Moon inertia matrix frame (PA) based on the diagonalization of the Moon matrix of inertia.} ILuRS adopts the PA frame as its natural choice, since it is directly defined by lunar ephemerides Euler angles, whereas ME requires additional averaging that can introduce misinterpretation \cite{Rambaux2026}. As PA and ME are equivalent in stability and accuracy, the WG prioritized robustness and accessibility, committing to provide transformation parameters (Helmert) between ILuRF and alternative frames (see Section~\ref{sec:helmert} and  and Fig.~\ref{fig:ILuRS_workflow}), ensuring straightforward updates with future ephemeris improvements.

\textcolor{black}{ILuRS co-rotates with the Moon and is anchored to an \textcolor{black}{undistorted} reference Moon — a geometrically ideal, undeformed body defined in the principal axis (PA) frame — to which tidal, rotational, and elastic deformations are subsequently added as corrections.} 

The International Lunar Reference frame (ILuRF) is based upon on lunar ephemerides and Moon gravity field, that provide the definitions of ILuRF origin, orientation and time-scale \cite{IAU2024} (see section \ref{sec:ILuRS}). The ILuRF materialization  is done with the coordinates of the Lunar Laser Ranging Retro-Reflectors (LLRRR) relative to the Moon's center of mass (CoM). The extension of ILuRF will be based on additional LLRRR that should be installed in the coming years. A supplementary product is the Helmert transformation parameters from ILuRF PA to previously published reference frames, including reference MEs.

\section{International Lunar Reference Frame ILuRF}
\label{sec:ILuRF}
Most of the details of this section can be found in \cite{Sosnica2026}. Here, we summarize only some important aspects.


\noindent \textbf{Context}

The definition of ILuRS and of its realization ILuRF is highly dependant on planetary and lunar ephemerides. 
In 2025, there are 3 state-of-the-art (SOA) ephemerides with similar accuracies : DE430\cite{Folkner2014}, INPOP21a \cite{2021NSTIM.110.....F} and EPM2021 \cite{Pitjeva2014}.
Comparisons between these 3 solutions can be retrieved for example in \cite{fienga24}.  
These 3 ephemerides were selected for the construction of ILuRS first realization, ILuRF2026.\\
 
\noindent \textbf{Method}

To avoid relying on a single ephemeris provider, \textcolor{black}{ILuRF adopts a weighted-averaging combination approach close to that of \cite{Beutler1995} for Global Navigation Satellite System  orbits}. Following \cite{Zajdel2025}, we combine SOA lunar ephemerides using a simplified Variance Component Estimator (VCE) to weight the Moon's positions, velocities, and orientation angles (in PA frame) consistently. The normalized weights are 0.451454072137127 for EPM2021,  0.380949596178373 for INPOP21a, and 0.167596331684500 for DE430. \textcolor{black}{The same weights are taken for the origin and the orientation parameters of the combinations.}
This method yields a single standardized ILuRF, with future updates expected as improved ephemerides become available. It also helps identify outliers, as illustrated by \cite{Sosnica2026} regarding DE440. The combined parameters (available on the website in Barycentric Coordinate Time TCB  - and Lunar Coordinate Time TCL-compatible versions) include:
\begin{itemize}
\item  ILuRF2026 origin: Moon CoM positions and velocities relative to Earth in Barycentric Celectial Reference System BCRS (in Barycentric Dynamical Time TDB)  ;\\
\item  ILuRF2026 orientation : Euler angles of Moon libration per \cite{Rambaux2026} (in TDB and TCL);\\
\item ILuRF2026 materialization : LLRRR coordinates  (in TDB and TCL).\\
\item ILuRF2026 time-scale: TCL-TCB computed at Moon CoM
\end{itemize}
Time series of  ILuRF2026 origin, orientation (so called LOOP for Lunar Origin and Orientation Parameters) and  TCL-TCB from 1970 to 2050 are provided with interpolation software (see Section~\ref{sec:distribution}). Additional LLR-related parameters are also delivered in TDB and TCL (Section~\ref{sec:distribution}). \\

\noindent \textbf{Definition of the Origin and Orientation}
\label{sec:loop}

\begin{figure}[h!]
   \centering
   \includegraphics[width=1.0\hsize]{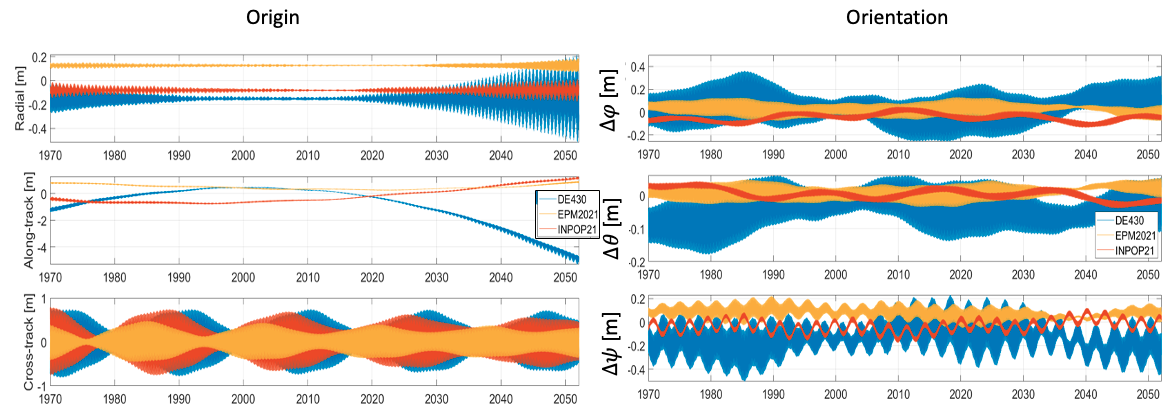}    
   \caption{ILuRF2026 origin realization (left-hand side) and orientation (right-hand side) differences between ILuRF and the contributing ephemerides, decomposed, on the left-hand side, into radial, along-track, and cross-track components and, on the right-hand side, into the 3 Euler angles ($\psi$, $\theta$, $\phi$) . Extracted from \cite{Sosnica2026}.}
         \label{fig:oo}
   \end{figure}

Fig. \ref{fig:oo} compares SOA ephemerides with ILuRF2026 in origin (left) and orientation (right). For origin, along-track differences reach up to 4 m in the prediction period, cross-track is stable at $\pm$1 m (well-defined orbital plane), and radial differences are at centimeter level, consistent with LLR constraints. For orientation, precession angle $\phi$ shows the largest deviation, with DE430 differences systematically larger due to its older data fit (up to 2014, versus 2020 for INPOP21a and EPM2021).
Internal ILuRF uncertainties \textcolor{black}{which are the VCE Weighted Mean Square Error (WMSE)\footnote{combination of the VCE weights $w$ and the differences between the VCE origins and orientations, $\hat{y}$ and those from SOA, $y$ such as WMSE=$\sum w_i (y_i-\hat{y}_i)/\sum w_i $}} are 17.6 cm over 2010–2030 (origin 15.3 cm, orientation 8.6 cm) and 31.6 cm over the full period \cite{Sosnica2026}. These errors stem mainly from the northern-hemisphere LLRRR distribution and limited Earth–Moon geometry, which also explain the larger along‑/cross‑track differences in Fig. \ref{fig:oo}. \\

\noindent \textbf{Validation with LLR}
\label{sec:llr}

\textcolor{black}{ILuRF2026 was validated against LLR data using multiple independent software packages, fixing its LOOP, LLRRR coordinates, Love number \(h_2\), and empirical corrections to compute residuals without adjusting other parameters. As shown in~\cite{Sosnica2026} and Table~\ref{tab:llr}, ILuRF2026 residuals are consistent with those of other ephemerides, confirming its robustness. Notably, ILuRF RMS closely matches EPM2021 (which received the highest weight in the combination), while DE430 shows larger RMS due to its older data fit (ending 2012 vs.~2020/2021 for INPOP21a/EPM2021). Independent groups are currently working on reproducing these findings with success. Preliminary NGLR-1 data (since March 2025) are promising, but its northern-hemisphere location limits the geometric gain for LOOP definition.} \\

\begin{table*}[ht]
\centering
\caption{RMS of LLR one-way residuals with kinematic corrections for selected high-performing stations in 1984-2025.}

\begin{tabular}{lcccccc}
\hline\hline
RMS [cm]  & Grasse (g) & Grasse (IR) & APOLLO & Wettzell & Matera & McDonald \\
\hline
DE430     & 3.0 & 3.9 & 3.3 & 5.0 & 4.3 & 3.9 \\
INPOP21a  & 1.9 & 2.2 & 2.5 & 2.0 & 3.6 & 4.3\\
EPM2021   & 1.7 & 1.8 & 2.4 & 2.6 & 3.3 & 3.6\\
ILuRF2026  \cite{Sosnica2026}    & 1.7 & 1.8 & 2.3 & 2.5 & 3.3 & 3.6\\
\hline\hline
\end{tabular}
\label{tab:llr}
\end{table*}

%

\noindent \textbf{Transformation to other reference frames}
\label{sec:helmert}

As given on the ILuRF2026 webiste and by \cite{Sosnica2026},  ILuRF2026 delivery includes transformation parameters to other reference frames, using 7 Helmert parameters (scale, translation, rotation) derived from fixed LLRRR control points \cite{Sosnica2026}. Transformations were computed for DE430, INPOP21a, and EPM2021 in PA, and for DE421 in ME—with both 7- and 3-parameter (no rotation) cases.  PA–PA errors are ~2–3 cm, while PA–ME DE421 errors reach 10 cm (7‑parameter) and 14 cm (3‑parameter); thus, the full 7‑parameter transformation is recommended for ME DE421. A significant correlation (0.97)  between X‑translation and Helmert scale is also noticed as an indication of the LLR geometry issue identified with the LOOP (see previous section). As demonstrated in \cite{Molli2026}, Moon southern hemisphere Very Long Baseline Interferometry  (VLBI) station is the most efficient way to address this issue as it will improve by 75$\%$ the definition of the LOOP and consequently the transformation parameters to other reference frames.\\


\noindent \textbf{Time-scale TCL}

\textcolor{black}{For the time-scale computations, we used on own software TEMPUS, an open-source Python library allows to compute coordinates times for planets and satellites of the solar system relative to TCB and TCG following the IAU 2000 \cite{Soffel2003} and 2024 resolutions \cite{IAU2024}. It has been benchmarked by comparison with published time-series for TCL-TCB such as \cite{klioner2025} and \cite{Lu2025} and for the Terrestrial Geocentric Time (TCG), TCG-TCB.
 As one see on Table \ref{tab:TCL-TCB}, the differences between these independent computations are below $5 \times 10^{-17}$ drift and 65 ps for the remaining detrended residuals, smaller than the present clock stability and accuracy.  A specific publication is under preparation for the presentation of the software and its benchmark.}
 
 \textcolor{black}{ILuRS being a geocentric system, TCL associated to ILuRS cannot be computed directly but by an averaged weighting of the TCL calculated using the 3 SOA ephemerides with the same weights as the one used for LOOP definition. 
As such TCLs are not delivered by the ephemeris groups, we use TEMPUS, for computing these TCL-TCB based on the 3 SOA dynamical modelings and their associated positions and velocities. But as one can see on Table \ref{tab:TCL-TCB}, the differences between the 3 SOA ephemerides do not introduce significant differences in their associated TCL-TCB as far as the current clock stability and accuracy are concerned.  
Consequently, the combined ILuRF2026 (TCL-TCB) also does not differ significantly from these estimations. 
Time-series of ILuRF2026 (TCL-TCB) are delivered on the ILuRS website and are accessible for the users. Are also delivered on the ILuRF2026 website the orientation angles in TCL. Once the TCB versus TDB scaling is applied, only the orientation rates are affected by the TCL-TCB time transformation introducing over 25 years a periodic trend of amplitude of about 2.7 mas on the precession angle and  an additional drift of 0.032 mas/day in
the rotation angle.\\
}
    \begin{table}
       \caption{Differences in TCL-TCB obtained with TEMPUS using the 3 SOA ephemerides and the ILuRF2026 combination over 200 years considering the contributions of the sun, the main planets, main belt asteroids and Kuiper belt objects. Columns 2 and 3 give the drift in s.s$^{-1}$ and maximum amplitudes of the detrended residuals  in picoseconds respectively. Are also given the differences between the TEMPUS TCL-TCB computation using the same modelings as \cite{klioner2025} and \cite{Lu2025} and the published time series.}
    \centering
\centering
    \begin{tabular}{l c c}
    \hline
  & drift & detrended \\
    & s.s$^{-1}$ & max amplitudes [ps]\\
    \hline
      TCL-TCB &  &  \\
    \hline
        \hline
     INPOP21a-DE430 & $1.73 \times 10^{-18}$ & $32.6$ \\
    INPOP21a-EPM2021 & $1.89 \times 10^{-18}$ & $64.5 $ \\
DE430-EPM2021 & $1.61 \times 10^{-19}$ & $11.2 $ \\
\hline
    INPOP21a-ILuRF2026 & $1.14 \times 10^{-18}$ & $33.6 $ \\
    EPM2021-ILuRF2026 & $ -7.48 \times 10^{-19}$ & $ 7.1 $ \\
DE430-ILuRF2026 & $ -5.86 \times 10^{-19}$ & $ 26.7 $ \\
\hline
\cite{klioner2025}-TEMPUS & $4 \times 10^{-20}$ & 6.0 \\
\cite{Lu2025}-TEMPUS &  $2 \times 10^{-17}$ & 40\\
\hline
    \hline
  TCG-TCB  &  &  \\
       \hline
\cite{2021NSTIM.110.....F}-TEMPUS & $1.70 \times 10^{-17}$ & 0.8 \\
\cite{klioner2025}-TEMPUS & $5.37 \times 10^{-21}$ & 6.0\\
\hline
    \end{tabular}
    \label{tab:TCL-TCB}
    \end{table}

\noindent \textbf{Delivery and associated parameters}
\label{sec:distribution}

The first ILuRS realisation ILuRF2026 is distributed via the websites \url{https://ilurs-6e772d.gitlab.io/} and  \url{http://ilurf.gssc.esa.int}. On this website, the users can find the time series for the ILuRF2026 LOOP (orientation and the origin) and TCL-TCB for the center of mass of the Moon, together with a Python library for the interpolation.  ILuRF2026 defining coordinates of the LLRRR are also provided together with auxiliary parameters necessary for the LLR validation as presented in Table \ref{tab:merged}.
Time series for Moon orientation, LLRRR coordinates, the Helmert transformation parameters and auxiliary parameters are provided in both TDB and TCL.

   \begin{table}
   \centering
       \caption{Auxiliary parameters necessary for the LLR validation and other applications in TDB. TCL-compatible values are provided on the ILuRS website}
           \label{tab:merged}
    \begin{minipage}{0.65\linewidth}
\centering
\begin{tabular}{lr}
\hline\hline
EMRAT &  81.300568541445642 \\
GM$_{EMB}$ &  8.9970113949394178E-10 AU$^3$/d$^2$\\
C$_{2,0}$ &  $-2.0321136949014574 \times 10^{-4}$ \\
R$_{moon}$ &  1738.0 km \\
\hline\hline
\end{tabular}

\end{minipage}
\begin{minipage}{0.30\linewidth}
\centering
\begin{tabular}{lr}
\hline\hline
$h_2$ &  0.0432 \\
$A_1$ &  4.4 mas\\
$A_2$ &  1.6 mas \\
$A_3$ &  1.2 mas \\
\hline\hline
\end{tabular}

\end{minipage}
    \end{table}

\section{Conclusions}

The current ILuRS realization, ILuRF2026, is accessible via the temporary website \url{https://ilurs-6e772d.gitlab.io/} and on the future \url{http://ilurf.gssc.esa.int}, along with auxiliary parameters, software, and the ILuRF2026 TCL time-scale (ESA GSSC site planned to open at fall 2026). ILuRS is proposed to become part of the the International Earth Rotation and Reference Systems  Product Center (accepted May 2026), though future inclusion criteria, update cadence, and governance remain to be defined.
Future improvements will come from new lunar surface observations, notably European Space Agency’s NovaMoon on the Argonaut missions (2029), which will add VLBI, radio transponders, and two Mini-RAFS clocks alongside LLRRR. As \cite{Molli2026} shows, VLBI in particular offers strong potential for refining ILuRF2026.

\subsubsection{Acknowledgements} The authors thank the members of the IAG JWG 1.1.3, and in particular K. Sosnica, D. Pavlov and N. Rambaux for their innovative work on ILuRF construction.

%
%
%
%

\end{document}